\documentclass[conference,10pt,a4paper]{IEEEtran}
\usepackage[pdftex]{graphicx}

\usepackage{url}       % Written by Donald Arseneau
\usepackage{amsmath}   % From the American Mathematical Society
\usepackage{array}
\begin{document}

% paper title
\title{Polish Grid Infrastructure for Science and Research}

% avoiding spaces at the end of the author lines is not a problem with
% conference papers because we don't use \thanks or \IEEEmembership

\author{\authorblockN{Ryszard Gokieli\authorrefmark{1},
Krzysztof Nawrocki\authorrefmark{11}, 
Adam Padee\authorrefmark{2}, 
Dorota Stojda\authorrefmark{3},
Karol Wawrzyniak\authorrefmark{4}\\ 
and Wojciech Wi\'slicki\authorrefmark{5}\authorrefmark{6}}
\authorblockA{\authorrefmark{1}A. So\l tan Institute for Nuclear Studies, Laboratory for High Energy Physics, Warsaw, Poland, \\ email: {\it gokieli@fuw.edu.pl}}
\authorblockA{\authorrefmark{11}A. So\l tan Institute for Nuclear Studies, Laboratory for High Energy Physics, Warsaw, Poland, \\ email: {\it nawrocki@fuw.edu.pl}}
\authorblockA{\authorrefmark{2}Warsaw University of Technology, Institute of Radioelectronics, Warsaw, Poland, \\ email: {\it apadee@ire.pw.edu.pl}}
\authorblockA{\authorrefmark{3}Copernicus Science Centre, Warsaw, Poland, email:{\it dorotas@icm.edu.pl}}
\authorblockA{\authorrefmark{4}University of Warsaw, Interdisciplinary Centre for Mathematical and Computational Modelling, Warsaw, Poland, \\ email: {\it kwawrzyn@icm.edu.pl}}
\authorblockA{\authorrefmark{5}University of Warsaw, Interdisciplinary Centre for Mathematical and Computational Modelling, Warsaw, Poland, \\ email: {\it wislicki@fuw.edu.pl}}
\authorblockA{\authorrefmark{6}A. So\l tan Institute for Nuclear Studies, Laboratory for High Energy Physics, Warsaw, Poland}
}

% use only for invited papers
%\specialpapernotice{(Invited Paper)}

% make the title area
\maketitle

\begin{abstract}
Structure, functionality, parameters and organization of the computing Grid in Poland is described, mainly from the perspective of high-energy particle physics community, currently its largest consumer and developer.
It represents distributed Tier-2 in the worldwide Grid infrastructure. 
It also provides services and resources for data-intensive applications in other sciences.
\end{abstract}


\section{Introduction}
Distributed computing was stimulated by human endeavours in the "big science" domain and, at the same time, by hopes of industry to commercialize developments in networking technologies. 
Some original ideas of organizing computing for science, as e.g. pioneering SETI@home \cite{setiathome}, initiated by D. Gedye for search for signals of extraterrestial civilizations, eventually evolved into coordinated networks supporting research projects of unprecedented scale, as e.g. the Large Hadron Collider (LHC\footnote{The LHC is a collider of protons at the total operational centre-of-mass energy of 14 TeV, located at the European Organization for Nuclear Research (CERN) near Geneva, Switzerland.}) Computing Grid (LCG) in particle physics \cite{lcg}.
\begin{figure}[h]
 \centering
 \includegraphics[width=8.3cm,height=7cm] {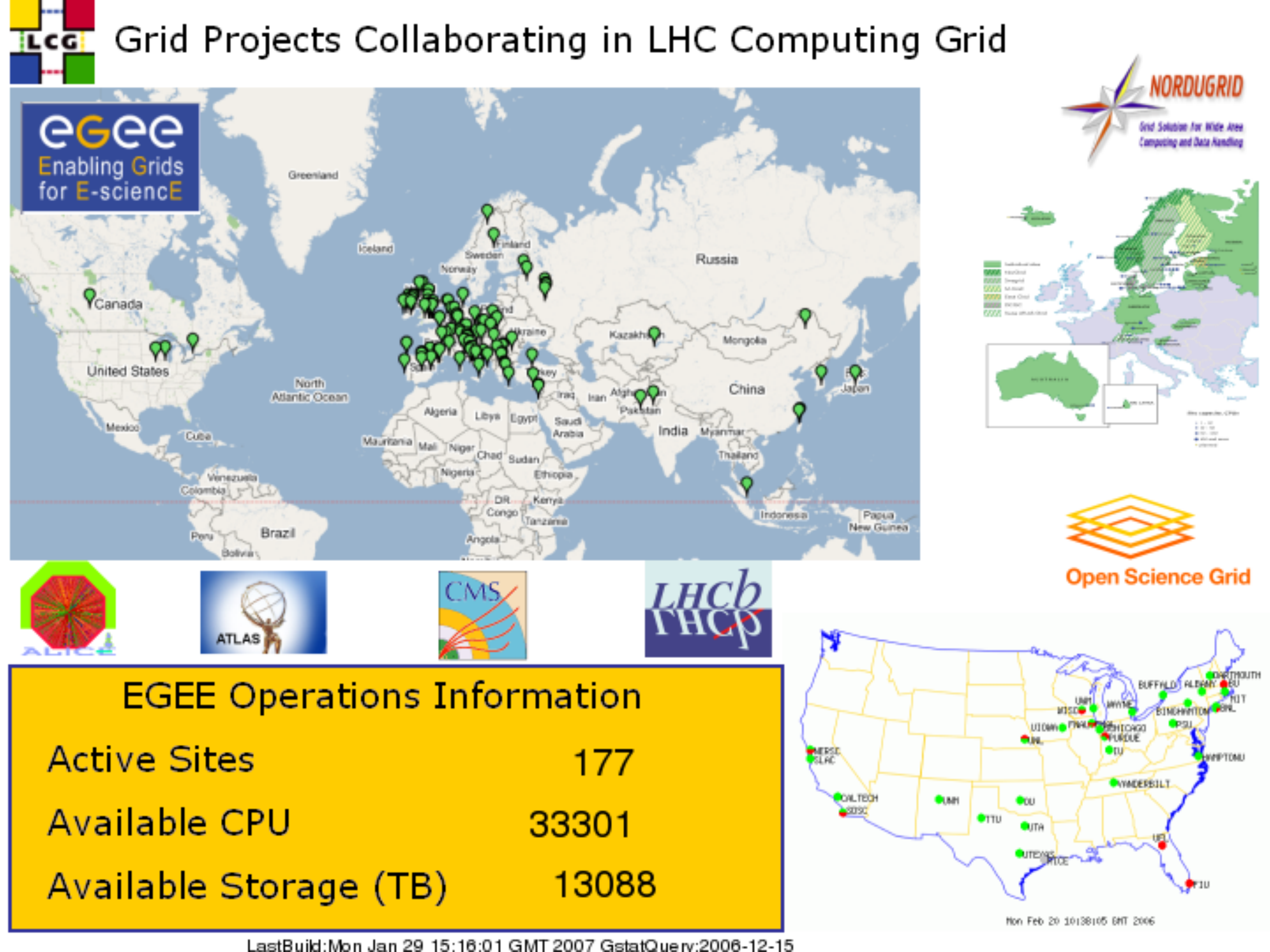}  
\caption{The map and basic data on the WLCG infrastructure. Collaborating sites are indicated as green points. Basic data for resources are given in the frame. The map is published on the LCG project pages \cite{lcg}.} \label{figure0}
\end{figure}
Nowadays, the Worlwide LHC Computing Grid (WLCG) is running on the infrastructure provided by the European initiative Enabling Grids for E-science (EGEE) \cite{egee1}, encompassing European national and regional Grids, coupled to the American Open Science Grid (OSG) and collaborating Grid centres in the Asia-Pacific region (cf. Fig. \ref{figure0}).
\begin{figure}[h]
 \centering
 \includegraphics[width=8.3cm,height=7cm] {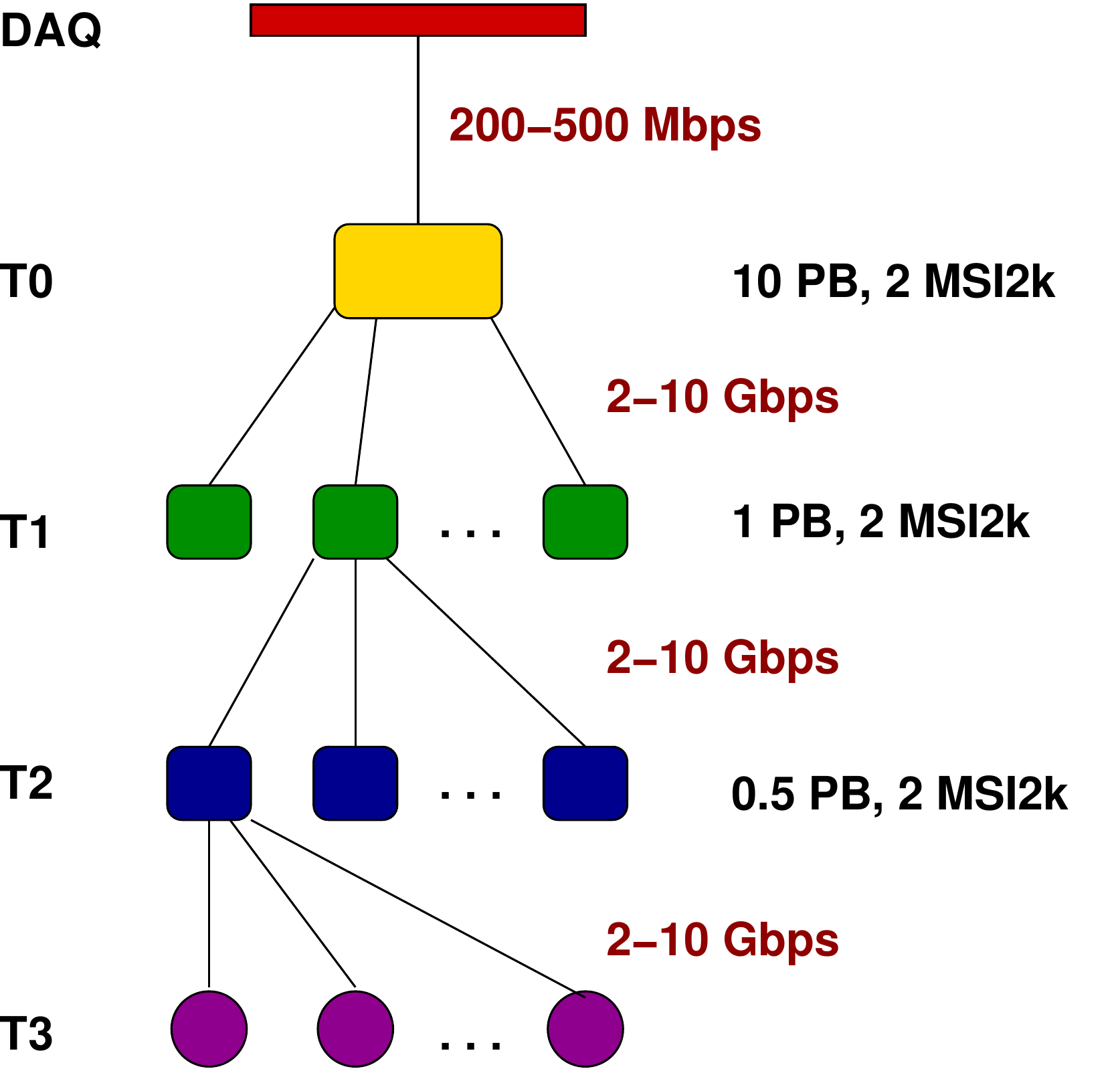}
\caption{Multitier tree-like LCG MONARC model. The DAQ stands for experiment's Data Acquisition System and T0-3 for Tier-0-3 levels of data processing. Required disc storage in petabytes (PB) and CPU in Mega SpecInt2000 (MSI2k), and connectivities between levels in bytes per second (bps) are indicated.} \label{figure1}
\end{figure}
This infrastructure is shared with a number of smaller projects.

\subsection{The Large-scale and Local Grid Architecture and Middleware}

Overall WLCG computing architecture is based on the hierarchical multi-tier model developed by MONARC Collaboration \cite{monarc}, as given in Fig. \ref{figure1}.
The top Tier-0 is responsible for storage of raw data coming from the experiment Data Acquisition System, its first off-line processing and distribution of data over Tier-1's. 
All data are copied to Tier-1 centres in order to speedup access during processing and ensure storage redundancy.
Data reprocessing and higher-level reconstructions of real and simulated data are foreseen to be performed at Tier-1 and Tier-2 levels.
Data and physics analyses are normally relegated to Tier-2 and -3 centres, closer to end-users. 
Tier-2s are powerful enough not only to support local needs but also to complement higher tiers with computing power and storage for more specialized purposes. Tier-2s are not required to provide with massive tape storage.

\begin{figure}[h]
 \centering
 \includegraphics[width=8.3cm,height=7cm] {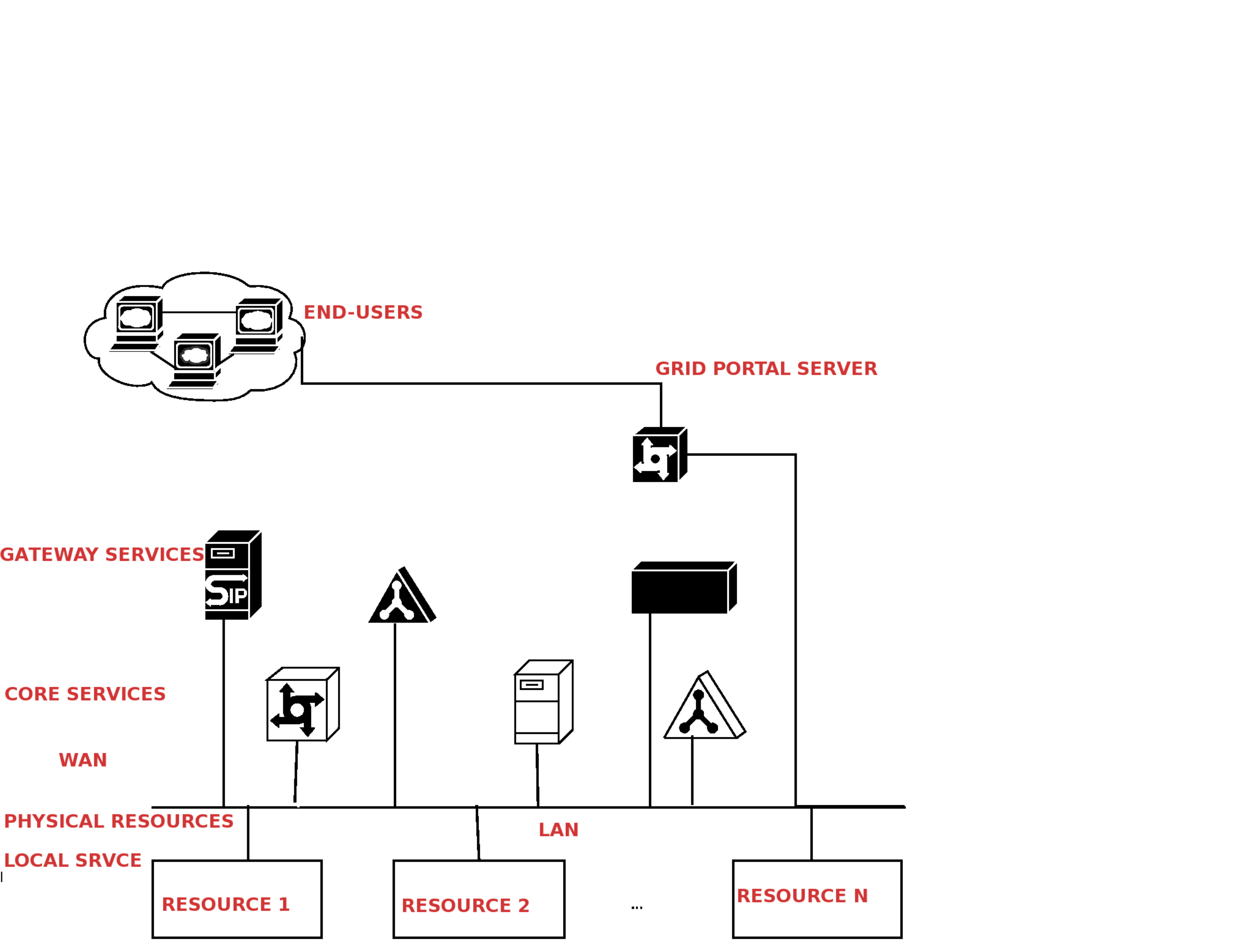}
\caption{Functional layers in Grid architecture. Physical resources and services in Local Area Networks (LANs) are available on the Wide Area Network (WAN), together with core and gateway services. End-users may access both the services and resources via dedicated portals ensuring easier workflow design.} \label{figure2}
\end{figure}

Based on the middleware gLite-3.0 \cite{laure1}, WLCG builds environment with the physical resource layer hidden behind core Grid services (cf. Fig. \ref{figure2}).  
Essential services, e.g. file and metadata catalogues, replica location, application resource catalogues or workflow engines, are either available directly for users or support other, complex services, e.g. Grid monitoring is used by resource brokers for process management.
Intelligent scheduling and resource brokering are now combined in a complex Workload Management service. 
This system comprises a set of Grid middleware components responsible for the distribution and management of tasks across the Grid.

\begin{figure}[h]
 \centering
 \includegraphics[width=8cm,height=3.5cm] {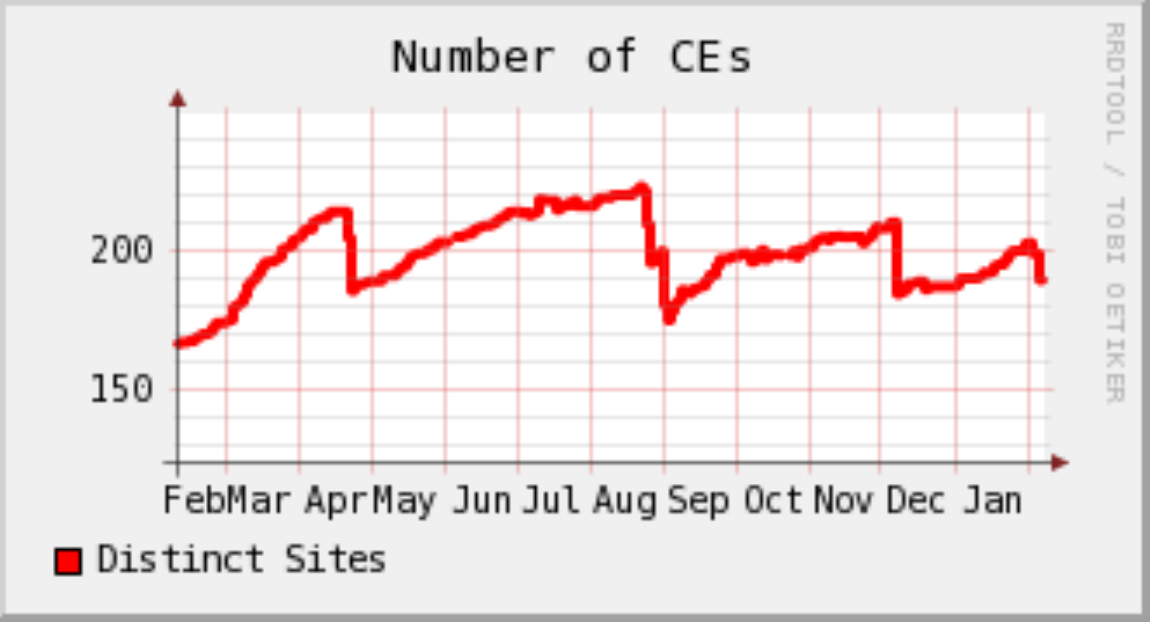}
 \includegraphics[width=8cm,height=3.5cm] {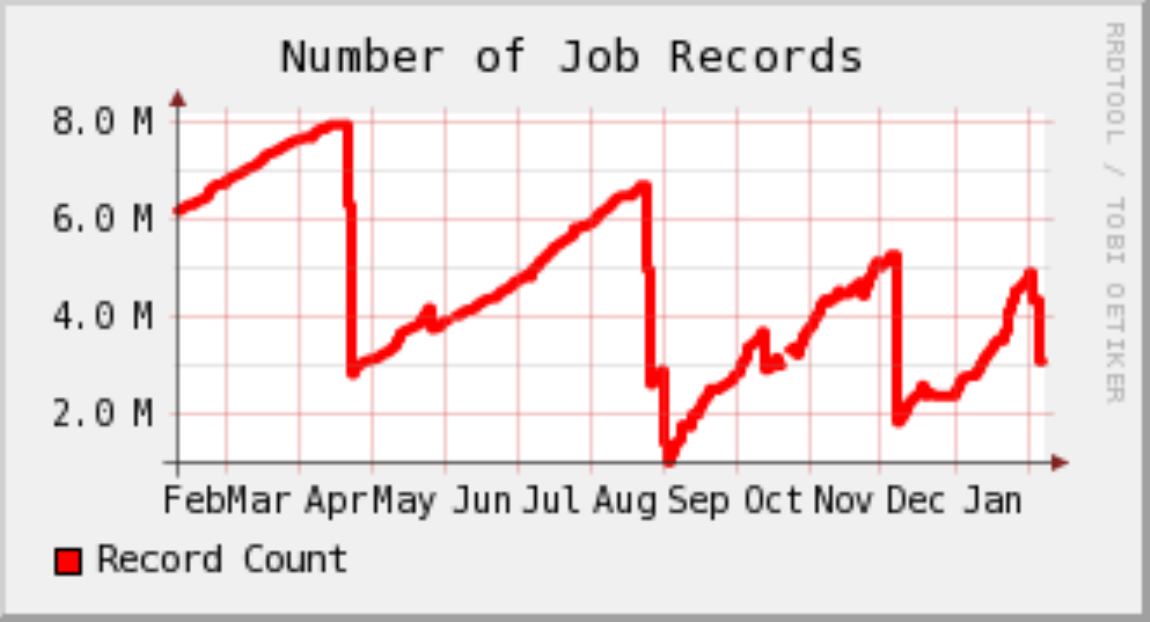}
\caption{Records of numbers of CEs (upper) and jobs (lower) monitored over a year. Drops are seen during Easter in April, summer holidays in August and Christmass in December.} \label{figure7}
\end{figure}

Information, monitoring and logging are available through the Relational Grid Monitoring Architecture (R-GMA) service, being an implementation of the GMA standard.
Fig.~\ref{figure7} presents counting of numbers of CEs and numbers of jobs monitored during one year, as provided by the R-GMA Monitoring Service.
Users interact with R-GMA through Application Programming Interfaces available for high-level programming languages.

Crucial for Grid management is the Distributed Grid Accounting System (DGAS).
Its purpose is to implement resource usage metering, accounting and account balancing in a fully distributed Grid environment.
The last function, i.e. account balancing, is still not in use, because the Grid has not yet come into commercial phase.

Large amounts of data are handled with DCache system \cite{dcache}.
Terabytes of data are distributed over many disc storage nodes but the name space is uniquely represented within a single file system tree. 
The system has shown to significantly improve the efficiency of connected tape storage systems, through caching, optimizing buffers and scheduled staging techniques. 
Furthermore, it optimizes the throughput to and from data clients as well as smoothing the load of the connected disc storage nodes by dynamically replicating datasets.

According to recent trend towards Service Oriented Architecture (SOA), Grid components are reengineered as Web-services and published on the net. 
Dedicated portals are used for designing complex workflows within SOA.

Users are organized in Virtual Organizations (VOs) and managed within VOs through the Virtual Organization Membership Service (VOMS).

\begin{figure}[h]
 \centering
 \includegraphics[width=8.3cm,height=6cm] {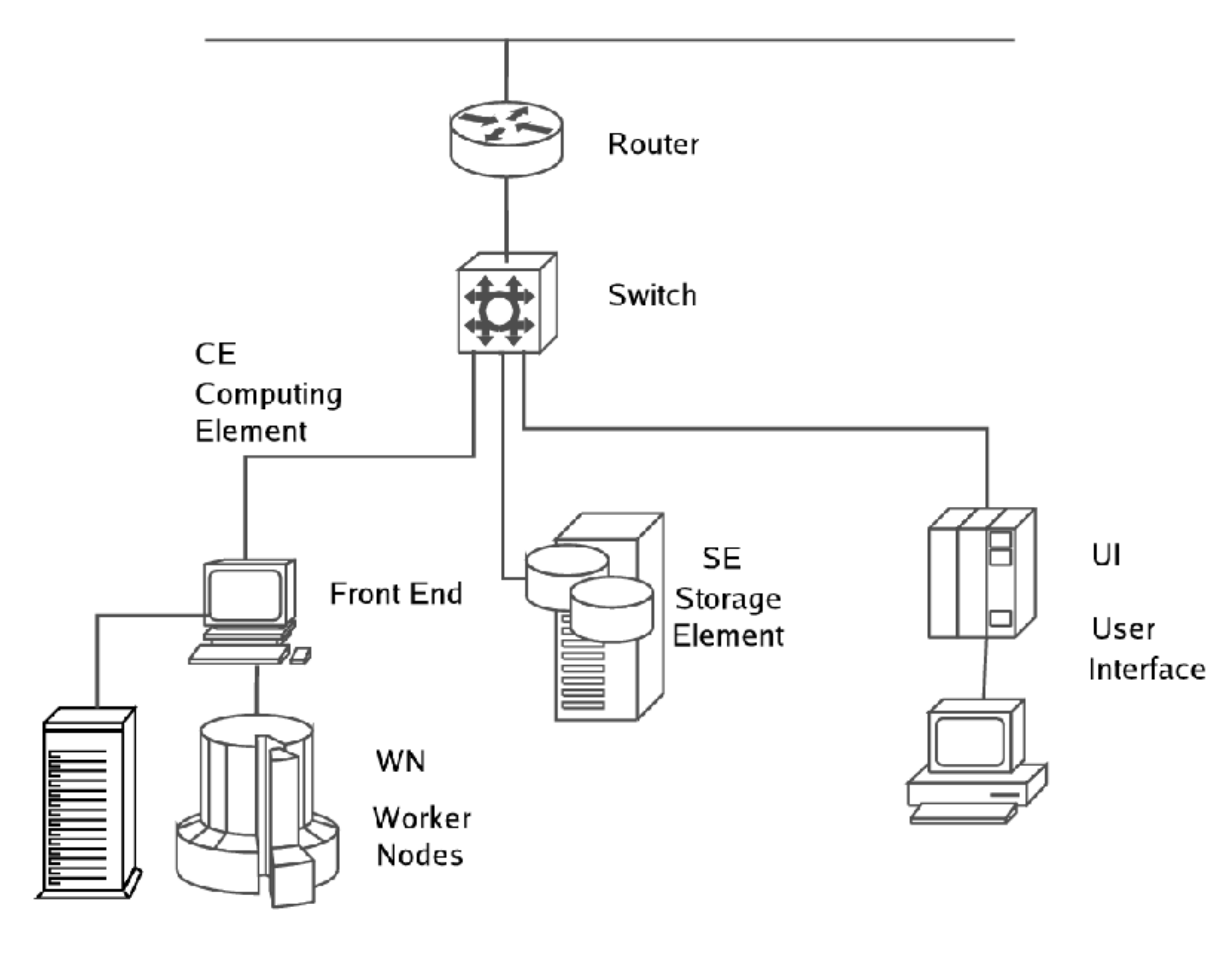}
\caption{Minimal site structure in computing Grid based on GLOBUS Toolkit. The CE and SE are on public IP numbers and can be accessed directly from WAN} \label{figure4}
\end{figure}

Each site is designed in a way typical for GLOBUS-operated \cite{globusalliance} Grids (cf. Fig. \ref{figure4}).
It contains the Computing Element (CE), normally consisting of the front-end machine playing gatekeeping role and a set of Worker Nodes (WNs).
The Storage Element (SE) is a main data container in a site and usually consists of disc matrices managed by dedicated machine. 
The User Interfaces (UIs), enabling user access to the infrastructure, may be either located on the spot or installed remotely. 

\subsection{Connectivity}

\begin{figure}[h]
 \centering
 \includegraphics[width=8.3cm,height=8cm] {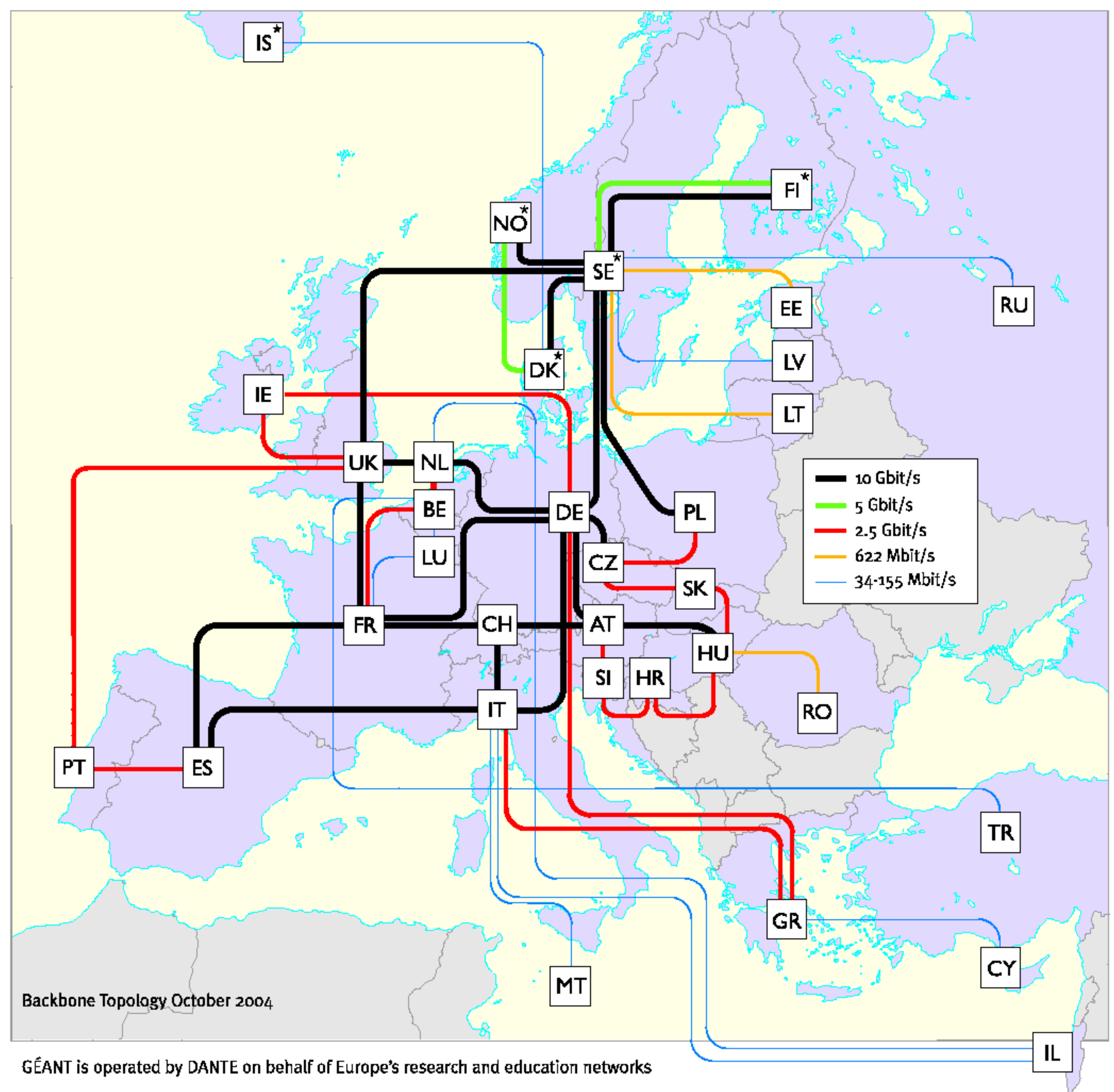}
\caption{G\'EANT connectivity scheme in Europe with bandwidths, indicated in colours, between national access points. The map is being updated on the G\'EANT project pages \cite{geant}.} \label{figure3}
\end{figure}

The Grids of EGEE and WLCG are built on top of the G\'EANT network \cite{geant} -- a collaboration of 26 national and research networks in Europe, led by the DANTE Company\footnote{DANTE is an acronym for {\it Delivery of Advanced Network Technology to Europe Limited}, located at Cambridge, England.}.
G\'EANT Project aims to deliver a quality-of-service gigabit speed backbone network for research in Europe. 
The G\'EANT connectivity scheme is given in Fig.~\ref{figure3}.
G\'EANT has 12~Gbps connectivity to North America and 2.5~Gbps to Japan and to Trans-Eurasia Information Network (TREIN2), thus ensuring collaboration of EGEE with OSG, Japaneese National Research Grid Initiative (NAREGI) and Asian-Pacific Grids. 
The entry point to Polish National Regional Network has bandwidth of 10 Gbps with 4470 B maximum transition unit on the switch.

\subsection{Computing Models of Principal End-users}

Four LHC experiments: ALICE \cite{alice1}, ATLAS \cite{atlas1}, CMS \cite{cms1} and LHCB \cite{lhcb1}, represent the largest consumers of resources on the Grid.
Raw data (RAW) coming from real experiment's DAQ or Monte Carlo simulation are recorded and processed off-line by reconstruction programs giving Event Summary Data (ESD), programs extracting physical variables and providing with Analysis Object Data (AOD), and to further data reduction, selection and filtering, leading to Event Tags (TAG).
In addition, derived streams of filtered data at the ESD and AOD levels and specialized data for detector alignment and calibration are recorded and analysed.
Grid is also used for the quasi on-line processing of data used for on-line calibration and filtering of data in the framework of the Interactive European Grid (IEG) project \cite{atlas2}.
Summary of data-flow parameters and requirements for off-line resources is given in Tab.~\ref{table1}.
\begin{table}[h]
\begin{center} \caption{Data flow and resource requirements for LHC experiments}
\label{table1}
\begin{tabular}{|c|c|c|c|c|}
 \hline
 & ALICE & ATLAS & CMS & LHCB \\ \hline
event rate (Hz) & 50 & 100 & 100 & 200 \\
Byte flow (MB/s) & 1250 & 100 & 100& 25 \\
CPU (kSI2k) & 30 & 23 & 40 & 13 \\
Storage (PB/y) & 25 & 17 & 30 & 7 \\ \hline
\end{tabular}
\end{center}
\end{table}
Such features of computing models as e.g. data flows, number of computing passes at each level, data redundancy, interactions of streams etc., differ between experiments according to specifics of physics processes and detectors to observe them.
Important differences in implementations of those models result in many experiment-specific tools and services used by each experimental group.

\section{Polish Tier-2 Infrastructure}

Polish physics groups are involved in four LHC experiments and in other high-rate experiments using the Grid, e.g. COMPASS \cite{compass1} at CERN and ZEUS \cite{zeus1} at Deutsches Elektronen-Synchrotron (DESY).
These groups are mostly affiliated at Cracow and Warsaw high-energy physics laboratories.

\subsection{Tier-2 -- Tier-1 Connectivity}

\begin{figure}[h]
 \centering
 \includegraphics[width=7.5cm,height=6.5cm] {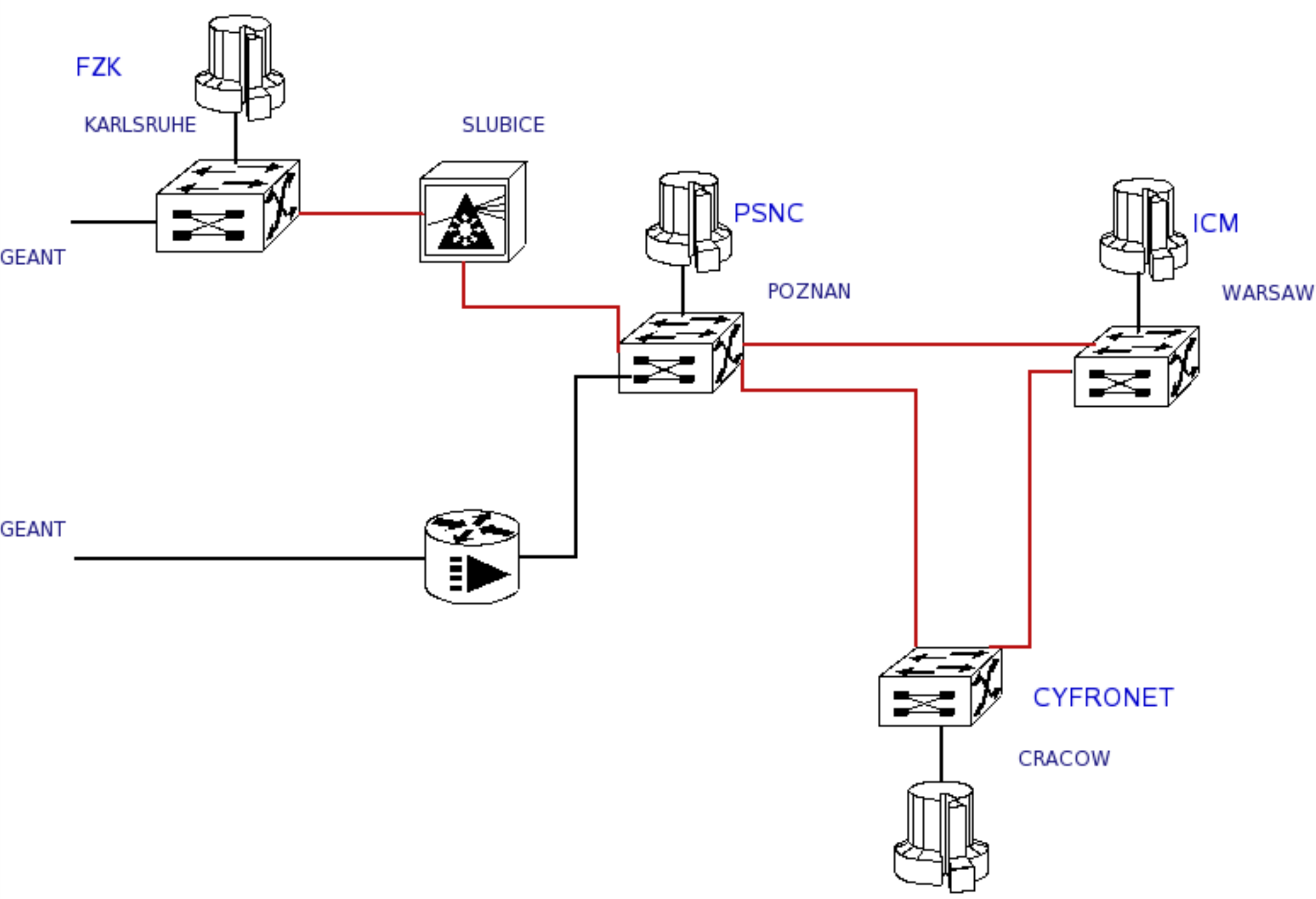} 
\caption{ Internal connectivity scheme for Polish distributed Tier-2 based on high-performance computing centres at Interdisciplinary Centre for Mathematical and Computational Modelling of Warsaw University (ICM), Cracow Academic Computing Centre of the Academy of Mining and Metallurgy (CYFRONET) and Poznan Supercomputing and Networking Centre of the Institute of Bioorganic Chemistry of Polish Academy of Sciences (PSNC). Polish national network is connected to German national network via wave-splitting DWDM multiplexer situated in S\l ubice.} \label{figure5}
\end{figure}

Large part of computing resources is located in these two cities and, in addition, in Pozna\'n, with no physics groups but where the operator of Polish backbone computing network PIONIER resides.   
These computing centres constitute Polish distributed Tier-2 connected to the Tier-1 centre at Forschungszentrum Karlsruhe (FZK) in Germany.
The PIONIER network interconnects Polish Tier-2 computing centres with a dedicated bandwidth of 1 Gbps and provides a bandwidth-splitting DWDM interface to the 10 Gbps backbone Deutsche Forschungsnetz (DFN) (cf. Fig. \ref{figure5}).
Polish Tier-2 centres and FZK constitute a Virtual Local Area Network (VLAN) with the address pool 212.191.227.xxx.

\subsection{Computing Infrastructure at Polish Tier-2 Centres}

Computing resources provided by Polish centres to Tier-2 are summarized in Tab.~\ref{table2}.
\begin{table}[h]
\begin{center} \caption{Physical computing resources provided by three Tier-2 computing Centres in Poland}
\label{table2}
\begin{tabular}{|c|c|c|}
 \hline
 site name & CPU available & Storage on SE (TB) \\ \hline
 AMD64.PSNC.PL & 222 & 4.3 \\
 CYFRONET-IA64 & 34 & 0.3 \\ 
 CYFRONET-LCG2 & 274 & 21.3 \\
 egee.man.poznan.pl & 132 & 5.2 \\
 WARSAW-EGEE & 224 & 5.9 \\ \hline
 Total & 886 & 37 \\ \hline
\end{tabular}
\end{center}
\end{table}
The clusters are not homogenous and different computing platforms and server hardware solutions are used.
As for the CPU, the 64-bit processor architecture prevails.
For example, at ICM the CE is based on AMD Opteron 250 processors assembled in Sun Fire v20 and v40 servers.
Rack-mounted WNs and StorEdge SE are interconnected with routable Nortel Baystack 5510-48T and Nortel Baystack 425-24T switches.
An automated IPMI tool was developed for efficient cluster management \cite{ipmi}. 

\begin{figure}[h]
 \centering
 \includegraphics[width=8cm,height=9cm] {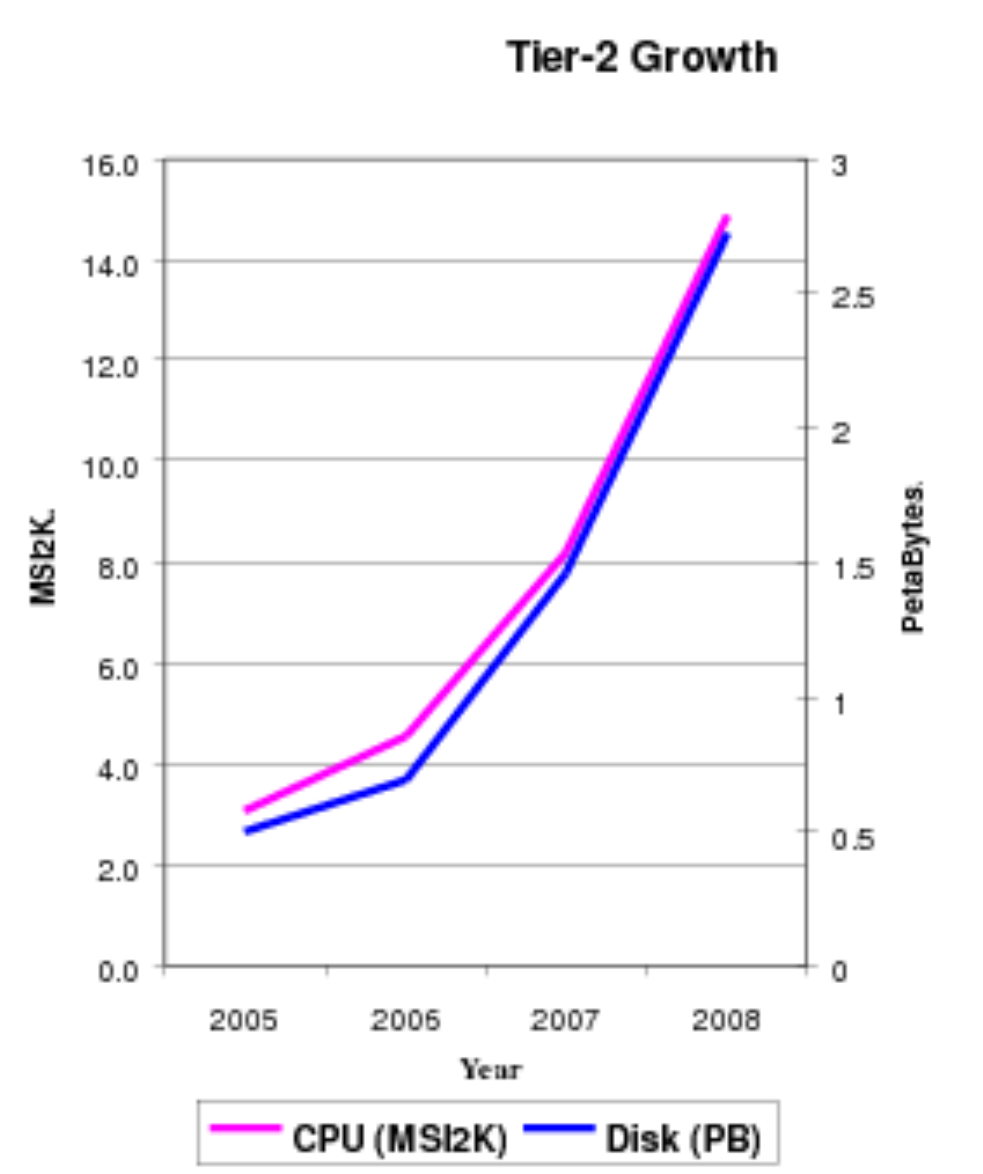}
\caption{Foreseen growth of CPU (purple) and disc storage (blue) in Tier-2. }\label{figure10}
\end{figure}

Just before start-up of the LHC Collider, Polish Tier-2 resources amount to almost 900 CPU and 37 TB of disc space on storage elements.
This represents 2.5 \% and 0.3\% of the total resources being 36,000 CPUs and 13.5~PB storage.
Similar figures for many Tier-2s are still below requirements mandatory at the LHC running time.
Depending on the experiment and its computing model, the CPU and disc storage are expected to be higher a couple of times and an order of magnitude, respectively.
Resource doubling every year, i.e. faster than the Moore's Law, is planned during LHC operation time (cf. Fig.~\ref{figure10}). 
Suitable investment for Polish Tier-2 is underway.

\begin{figure}[h]
 \centering
 \includegraphics[width=8cm,height=3.5cm] {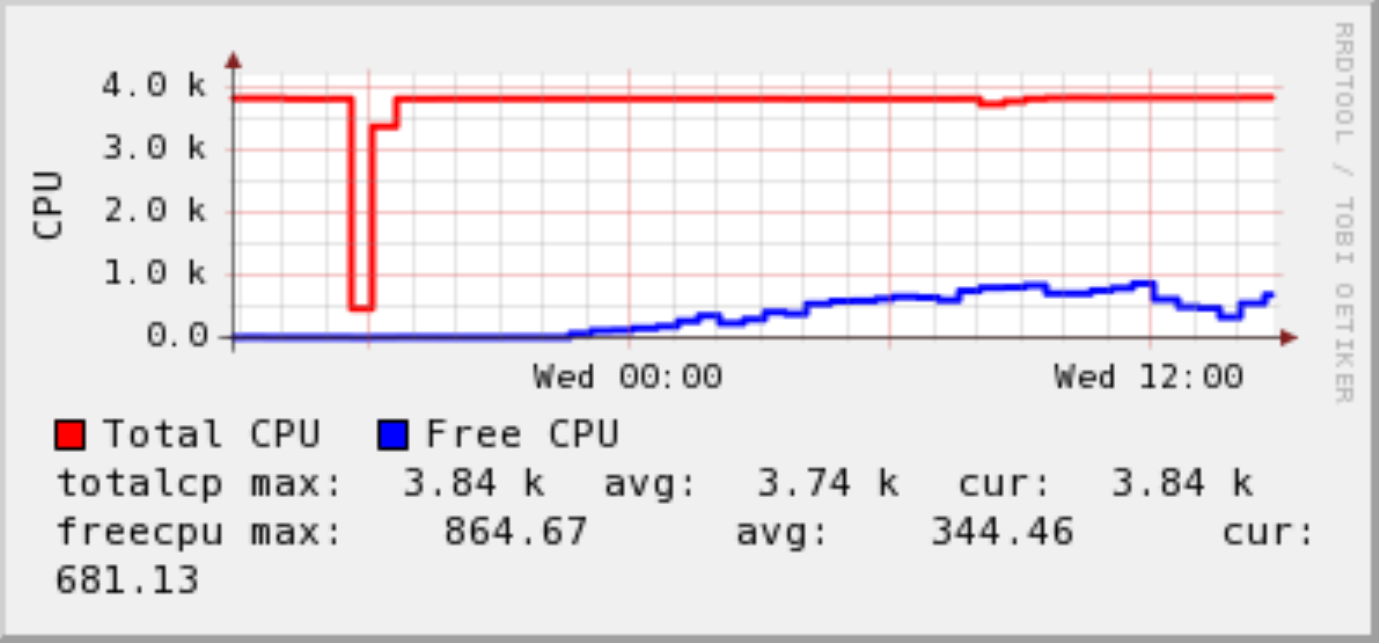}
 \includegraphics[width=8cm,height=3.5cm] {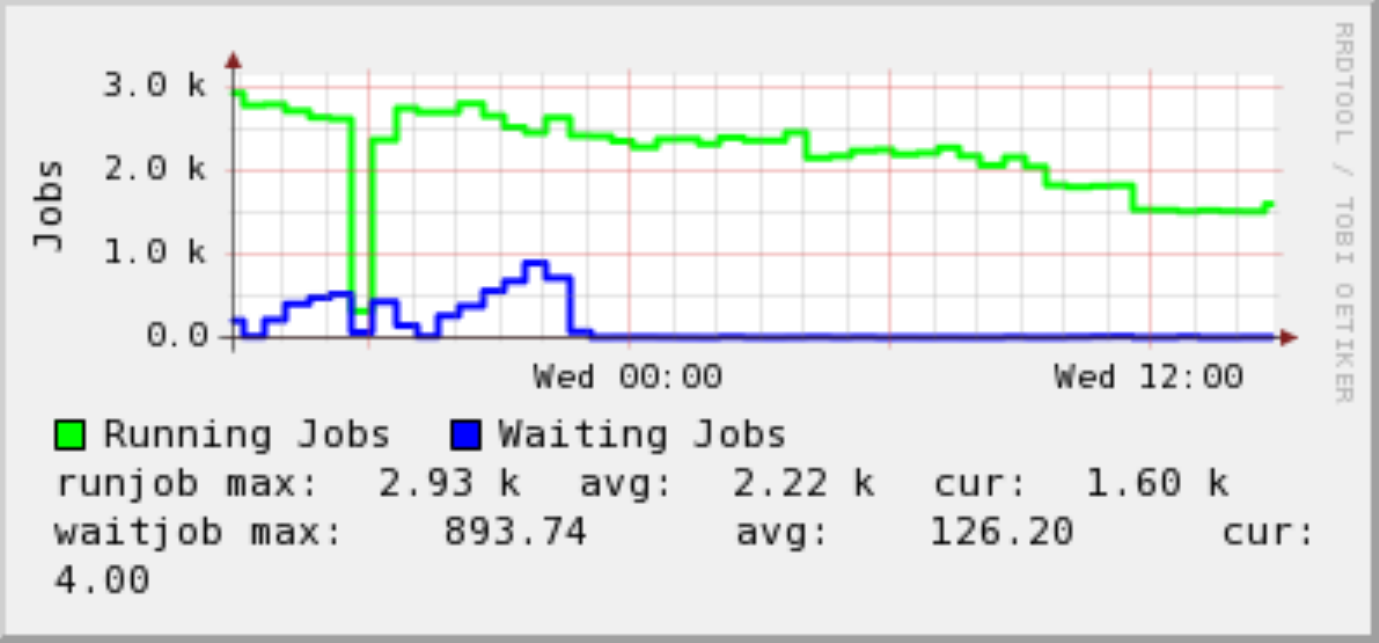}
 \includegraphics[width=8cm,height=3.5cm] {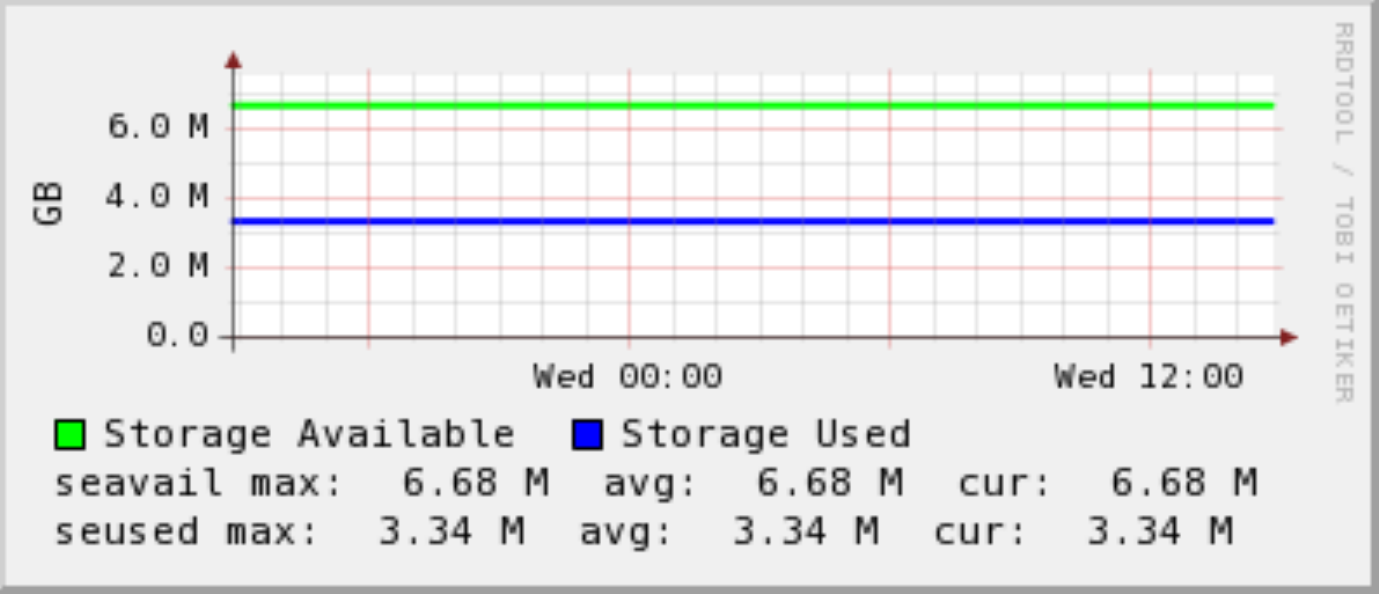}
\caption{Monitoring plots provided by Grid Index Information Service tool. In the upper panel, total number of Working Nodes' CPUs available for users is shown in red and free CPUs in blue. The middle panel presents numbers of runnng (green) and waiting (blue) jobs. Disk storage on Storage Elements available (green) and used (blue). Temporary connectivity drop is seen as a dip in two upper figures.} \label{figure6}
\end{figure}

\begin{figure}[h]
 \centering
 \includegraphics[width=8cm,height=3.5cm] {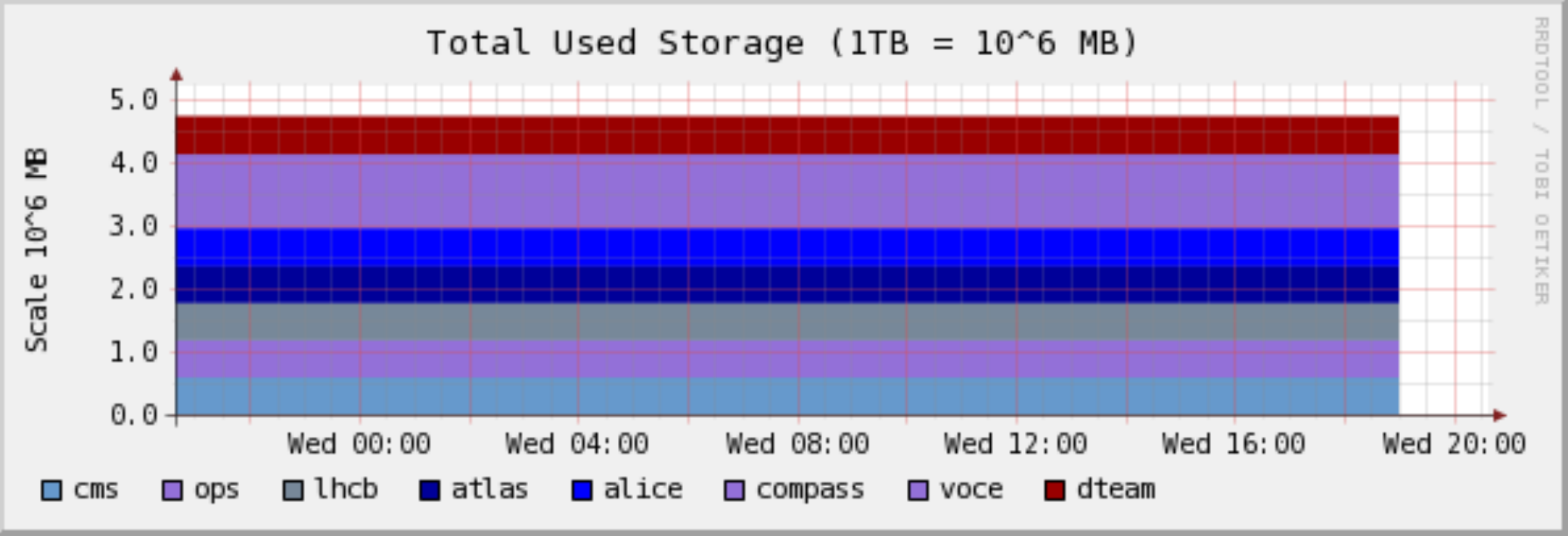}
 \includegraphics[width=8cm,height=3.5cm] {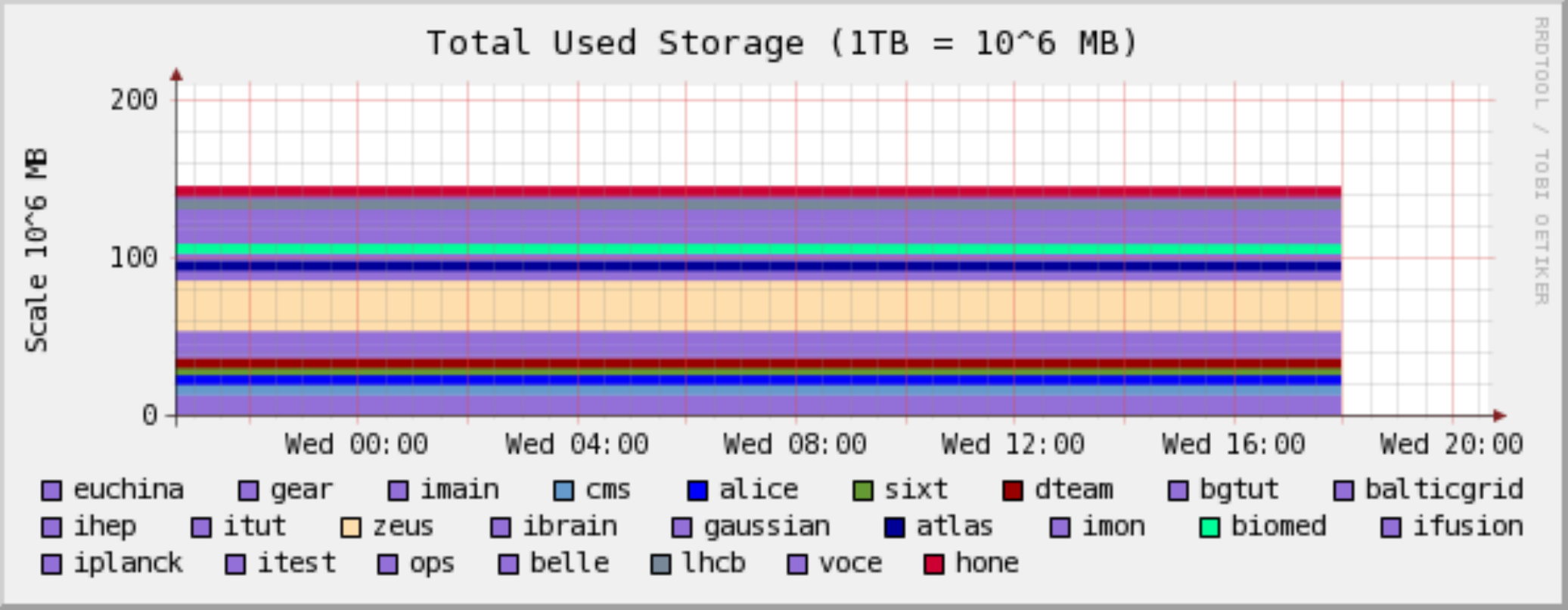}
\caption{Usage of disc storage by VOs at ICM (upper) and CYFRONET (lower).} \label{figure8}
\end{figure}

Each site on the Grid is permanently monitored by Grid Operational Centre and its status, both physical and functional, is made available on the Academia Sinica Web Host \cite{goc1}.
Example plots showing actual numbers of CPU usage, numbers of jobs and disc storage usage are given in Fig.~\ref{figure6}.
More detailed insight into disc storage distribution over VOs is displayed in Fig.~\ref{figure8}.

\subsection{VO Support and Resource Sharing}
\begin{table}[h]
\begin{center} \caption{Virtual Organizations supported by Polish Tier-2
 centres}
\label{table3}
\begin{tabular}{|c|c|c|}
 \hline
 Site & VO & Normalized CPU share $(\%)$ \\ \hline
AMD64.PSNC.PL & ATLAS & 73.9 \\
              & BALTICGRID & 6.9  \\
              & CMS & 4.1 \\
              & COMPCHEM & 11.5 \\
              & DTEAM & 0.2 \\
              & OPS & 0.1 \\
              & VOCE & 3.3 \\ \hline
CYFRONET-IA64 & ALICE & 14.3 \\
              & ATLAS & 26.8 \\
              & BALTICGRID & 6.6 \\
              & BIOMED & 2.0 \\
              & COMPCHEM & 25.2 \\
              & DTEAM & 0.3 \\
              & EUCHINA & 18.9 \\
              & LHCB & 0.4 \\
              & OPS & 0.1 \\
              & VOCE & 5.5 \\ \hline
CYFRONET-LCG2 & ATLAS & 26.7 \\
              & BALTICGRID & 0.5 \\
              & BIOMED & 30.2 \\
              & CMS & 11.3 \\
              & DTEAM & 0. \\
              & EUCHINA & 1.0\\
              & GEAR & 0.1 \\
              & HONE & 1.2 \\
              & LHCB & 26.1 \\
              & OPS & 0. \\
              & VOCE & 0.8 \\
              & ZEUS & 2.27 \\ \hline
egee.man.poznan.pl & ALICE & 0. \\
                   & ATLAS & 83.3 \\
                   & BALTICGRID & 1.6 \\
                   & CMS & 0.1 \\
                   & COMPCHEM & 12.30 \\
                   & DTEAM & 0.1 \\
                   & OPS & 0. \\
                   & VOCE & 2.6 \\ \hline
WARSAW-EGEE & ATLAS & 11.6 \\
            & CMS & 54.1 \\
            & COMPASS & 0.1 \\
            & DTEAM & 0.1 \\
            & LHCB & 32.0 \\
            & OPS & 0.\\
            & VOCE & 2.1 \\ \hline
\end{tabular}
\end{center}
\end{table}
Tab.~\ref{table3} shows resourse sharing over virtual organizations in Polish Tier-2 clusters.
Besides already mentioned VOs, one finds VOs related to the Baltic Grid project (BALTGRID), biology (BIOMED), chemistry (COMPCHEM), internal EGEE
 development VO (DTEAM), the EU-China Grid initiative (EUCHINA), Central
 European Federation VO (VOCE) and a couple of minor VOs.
In this report we do not distinguish between EGEE VOs, official global VOs, official local VOs and others, although these distinctions are important from managerial viewpoint.

Inspection of the table reveals differences in local policies of resource allocation to VOs.
There is no yet official regulation for these policies and resource allocations are usually negotiated between user communities and site managements.
In order to ensure optimal CPU usage, fair share system is normally implemented in queues, unless given VO uses privately funded machines.
In future, the quality-of-service system allowing reservations and hiring is foreseen for both computational resources and communication bandwidth.
These issues are related to future commercialization of the Grid.

\section{Operations of Polish Tier-2}

\subsection{Daily Operations and User Support}

European Grid is supposed to provide a permanent and reliable infrastructure for research and science.
The hardware is run by staff of participating institutions and is under local responsibility.
Both central and local services are run by dedicated groups of people and are shared between partners, depending on their competence, size and needs of regional scientific groups.
The case of large LHC collaborations, consisting even of thousands of researchers, somewhat violates this scheme.
While the lowest-order Tier-3 nodes are traditionally situated in scientific institutes, Tier-2's and Tier-1's are often run by large regional or national computing centres, capable of fulfilling operational requirements.

Daily operations are monitored by Grid Operational Centre (GOC) located at UK \cite{goc2}.
The GOC is responsible for coordinating the overall operation of the Grid. 
It acts as a central point of operational information such as configuration information and contact details.
The GOC has responsibility for monitoring the operation of the Grid Infrastructure as a whole, devising and managing mechanisms and procedures which encourage optimal operation of the Grid, and working with Local Support Groups to assist them in providing the best possible service while their equipment is connected to the Grid.

Basic functionality of services is regularly tested and monitored using Site Functional Tests (SFTs). 
The SFT uses a small test job that runs at each site and determines the availability of the main Grid functions. 
Similarly, the Grid Status Monitor retrieves information published by each site about its status. 
Their use and subsequent triggering of follow-up action is supervised by the Core Infrastructure Centre (CIC) on Duty staff raising operational tickets against sites to resolve observed problems.

\begin{figure}[b]
 \centering
 \includegraphics[width=8cm,height=5cm] {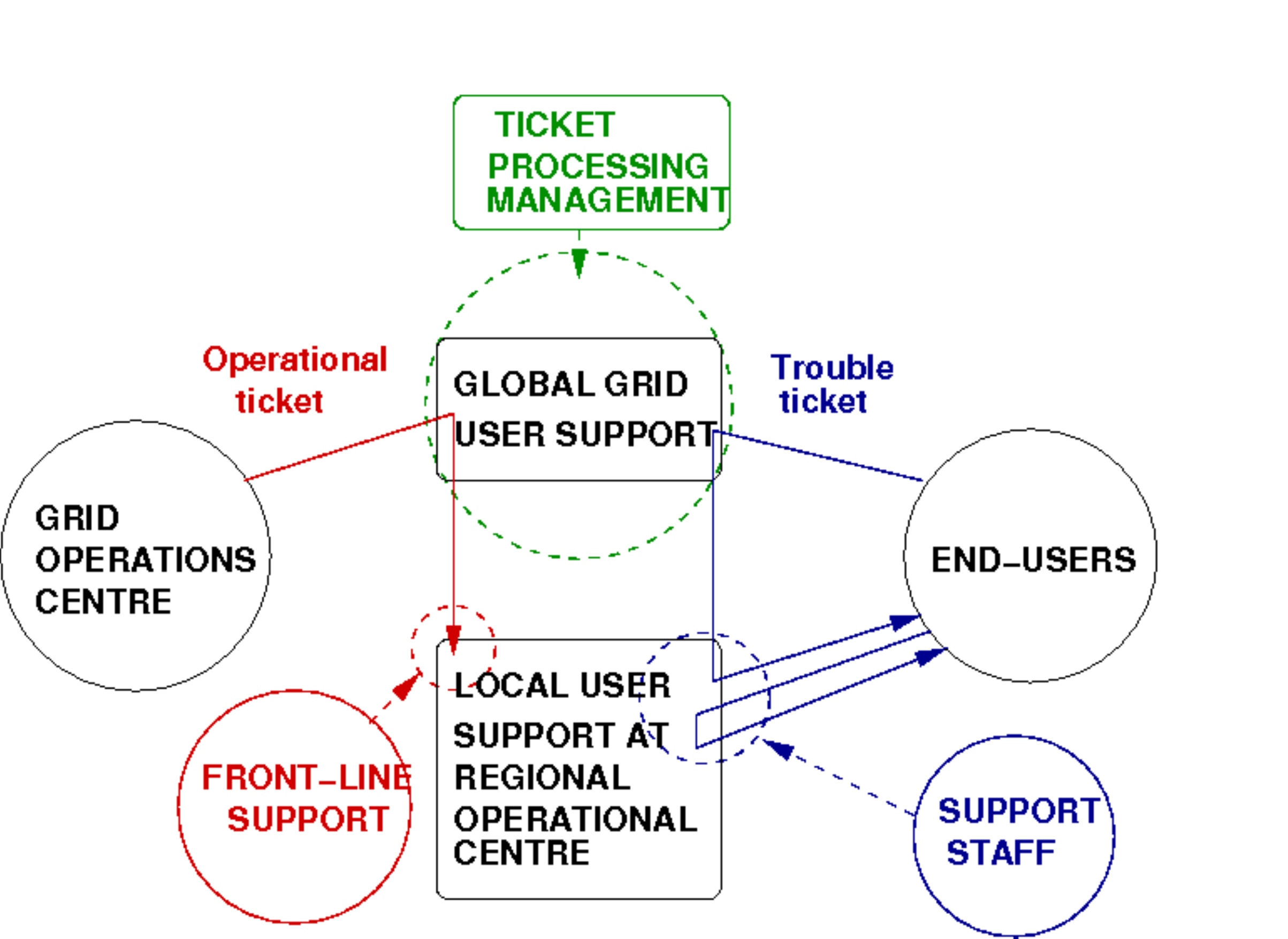}
\caption{Tickets flow scheme. Tickets issued by SFT failures are put onto GGUS and automatically directed to Local Support units and solved there. End-users' tickets can be either put to GGUS or to Local Support units. These tickets are solved either by local staff or by supporters from dedicated support groups, depending on the type of the problem. Daily ticket handling is performed by TPM group. Solved tickets are stored in a database.}\label{figure9}
\end{figure}

Two-level ticketing system is incorporated.
Ticket flow diagram is displayed in Fig.~\ref{figure9}.

Global Grid User Support (GGUS) portal run by FZK is a principal entry point for all sorts of tickets \cite{ggus}.
Daily ticket operations, including initial recognition of the type of the problem, opening and assigning ticket to supporters, directing it to Local Support units, care about timely solving and contact with users, is a duty of Ticket Processing Management Group (TPM). 
The TPM works in a shift system, 5 days a week, 8 hours a day.
SFT failure tickets are send to GGUS and redirected automatically to Local Support units and from there to the front-line supporters at sites.
Another type tickets are issued by end-users who may encounter many kinds of problems with a system or with applications.
User tickets, depending on the type of the problem, are either solved by dedicated group of supporters asked by TPM from GGUS, or redirected to Local Support units and solved there.
 
There is at least one Local Support unit in Federation.
For Polish Tier-2, ticketing tool \cite{helpdesk1} is running using {\it One-or-zero} portal software \cite{ooz}.

\section{Interactive Grid Infrastructure}

Bulk of applications in experimental particle physics and other sciences needs a high-throughput batch processing of large amounts of data.
For a number of applications, however, often interaction with intermediate results or fast response to a well defined computational problem is desired.
This sort of applications, called (quasi-) interactive, draws an attention of Grid community since almost beginning.
In Poland, involvement in deploying such applications on the Grid and building appropriate tools and infrastructure for them, dates back to the CrossGrid Project \cite{crossgrid1} and is nowadays continued in the framework of IEG project \cite{atlas2}.

The IEG resources are split into two separated infrastructures: 
\begin{itemize}
\item the production infrastructure aimed at providing computing and storage resources for the end-users running scientific applications, 
\item the development infrastructure being fully independent of the production and aimed at supporting the Project software development, the test of new middleware and its rollout process. As such this infrastructure does not provide a service as stable and reliable as the production. Development sites may also be occasionally reconfigured with specific setups to evaluate or validate software components.
\end{itemize}
Currently, the production infrastructure provides with 300 CPU cores and 8 TB disc storage, located in 8 computing centres in Europe, with a considerable contribution of three Polish computing centres.

The IEG supports the following interactive applications:
\begin{itemize}
\item Ultra Sound Computer Tomography.
\item Medical Applications on Brain Images
\item Flood Forecasting application. This application was first deployed on the CrossGrid testbed.
\item Visualisation of Baltic Wave Model.
\item Evolution of pollution clouds in the atmosphere. This application was first deployed on the CrossGrid testbed.
\item ATLAS online monitoring and calibration system. This application is related to ATLAS experiment at LHC but it does not run on the WLCG infrastructure. 
\item Analysis of Maps of Cosmic Microwave Background.
\item Visualization of Plasma in Fusion Reactors. This application runs also in less interactive mode on the EGEE infrastructure.
\end{itemize}

\section{Training, Demonstration and Diffusion Activities}

Being a global-scale initiative with large investment and social impact, computing Grid needs associated actions attracting and training users, and explaining the newest Grid technology to wider public.

Training is provided by organizing courses, normally given by staff members of academic partner institutes of large Grid projects (cf. e.g. refs \cite{atlas2,polgrid}), and using dedicated training infrastructure.
Courses are attended by students of universities at the engineering, M.Sc. and Ph.D. levels, practicing scientists of informatics, natural sciences and engineering, developers and managerial staff from commercial companies.

Dissemination of Grid technology is assured by its active promotion in communities of the actual and potential users by using press news and specialized publications, participating in conferences of possibly wide spectrum of subjects, and through media \cite{polgrid,atvn}.
The on-demand TV, being nowadays a distributor of knowledge, may shortly become Grid's customer exploiting its computing power for own programme casting and production.

\section{Outlook and Perspectives}

The WLCG is an example of well developed Grid infrastructure for science.
To large extent, however, it was designed for specific needs of experimental particle physics where high-throughput, massive. asynchronous data-intensive processing of segmented data is needed.
Parallel processing and using Massage Passing Interface software is rather rare.
Occasional usage of such gateway services like application resource catalogs or workflow engines is not a common practice.
Workflow management is quite often done semi-manually and resource brokerage is still far from being adaptive and autonomous.
Prospective line of development guides toward SOA where applications are distributed over the network and are accessible from everywhere as services.
Data are going to be virtualized (dCache) and workflows are dynamically designed and redesigned according to needs (cf. e.g. PGrade portal \cite{pgrade}).

After fulfilling LHC commitments, Polish Tier-2 should evolve towards new computational paradigm where complex reserach scenarios are executed in response to external events (e.g. rapid weather change) in closed loops with instruments and humans (interactivity). 
On-demand allocation of computing resources should ensure solving identified important problems.

From commercial perspective, other aspects of Grid should be underlined.
For pure research, robustness and security for applications is not really critical unless facilities are being built.
Scientific groups gladly relegate operations to commercial entities.
But this practice often results undesirably for science because business itself is only interested in research in case of visible income.
Interesting game between two aspects of distributed, large-scale computing: the economic and the research, is in front of us.

\end{document}